\documentclass[aps,prb,twocolumn,showpacs,preprintnumbers,amsmath,amssymb,superscriptaddress]{revtex4}%

\usepackage{graphicx}
\usepackage{dcolumn}
\usepackage{bm}
\usepackage{color}

\begin{document}

\preprint{PREPRINT (\today)}

\newpage

\title{Anisotropic superconducting properties of single-crystalline FeSe$_{0.5}$Te$_{0.5}$}

\author{M.~Bendele}
\email{markus.bendele@physik.uzh.ch}
\affiliation{Physik-Institut der Universit\"{a}t Z\"{u}rich, Winterthurerstrasse 190, CH-8057 Z\"{u}rich, Switzerland}
\affiliation{Laboratory for Muon Spin Spectroscopy, Paul Scherrer
Institute, CH-5232 Villigen PSI, Switzerland}

\author{S.~Weyeneth}
\affiliation{Physik-Institut der Universit\"{a}t Z\"{u}rich, Winterthurerstrasse 190, CH-8057 Z\"{u}rich, Switzerland}

\author{R.~Puzniak}
\affiliation{Institute of Physics, Polish Academy of Sciences, Aleja Lotnik\'ow 32/46, PL-02-668 Warsaw, Poland}

\author{A.~Maisuradze}
\affiliation{Laboratory for Muon Spin Spectroscopy, Paul Scherrer
Institute, CH-5232 Villigen PSI, Switzerland}

\author{E.~Pomjakushina}
\affiliation{Laboratory for Developments and Methods, Paul Scherrer
Institute, CH-5232 Villigen PSI, Switzerland}

\author{K.~Conder}
\affiliation{Laboratory for Developments and Methods, Paul Scherrer
Institute, CH-5232 Villigen PSI, Switzerland}

\author{V.~Pomjakushin}
\affiliation{Laboratory for Neutron Scattering, ETHZ \emph{\&} PSI, CH-5232 Villigen PSI, Switzerland}

\author{H.~Luetkens}
\affiliation{Laboratory for Muon Spin Spectroscopy, Paul Scherrer
Institute, CH-5232 Villigen PSI, Switzerland}

\author{S.~Katrych}
\affiliation{Laboratory for Solid State Physics, ETH Zurich, CH-8093 Z\"{u}rich, Switzerland}

\author{A.~Wisniewski}
\affiliation{Institute of Physics, Polish Academy of Sciences, Aleja Lotnik\'ow 32/46, PL-02-668 Warsaw, Poland}

\author{R.~Khasanov}
\affiliation{Laboratory for Muon Spin Spectroscopy, Paul Scherrer
Institute, CH-5232 Villigen PSI, Switzerland}

\author{H.~Keller}
\affiliation{Physik-Institut der Universit\"{a}t Z\"{u}rich, Winterthurerstrasse 190, CH-8057 Z\"{u}rich, Switzerland}

\begin{abstract}
Iron-chalcogenide single crystals with the nominal composition FeSe$_{0.5}$Te$_{0.5}$ and a transition temperature of $T_{c}\simeq14.6$~K were synthesized by the Bridgman method.
The structural and anisotropic superconducting properties of those crystals were investigated by means of single crystal X-ray and neutron powder diffraction, SQUID and torque magnetometry, and muon-spin rotation. Room temperature neutron powder diffraction reveals that 95\% of the crystal volume is of the same tetragonal structure as PbO.
The structure refinement yields a stoichiometry of Fe$_{1.045}$Se$_{0.406}$Te$_{0.594}$.
Additionally, a minor hexagonal Fe$_{7}$Se$_{8}$ impurity phase was identified.
The magnetic penetration depth $\lambda$ at zero temperature was found to be $\lambda_{ab}(0)=491(8)$~nm in the $ab$-plane and $\lambda_{c}(0)=1320(14)$~nm along the $c$-axis.
The zero-temperature value of the superfluid density $\rho_s(0)\propto \lambda^{-2}(0)$ obeys the empirical Uemura relation observed for various unconventional superconductors, including cuprates and iron-pnictides.
The temperature dependences of both $\lambda_{ab}$ and $\lambda_c$ are well described by a two-gap $s$+$s$-wave model with the zero-temperature gap values of $\Delta_S(0)=0.51(3)$~meV and $\Delta_L(0)=2.61(9)$~meV for the small and the large gap, respectively.
The magnetic penetration depth anisotropy parameter $\gamma_\lambda(T)$ = $\lambda_{c}(T)$/$\lambda_{ab}(T)$ increases with decreasing temperature, in agreement with $\gamma_\lambda(T)$ observed in the iron-pnictide superconductors.
\end{abstract}


\maketitle

\section{Introduction}
Since the discovery of superconductivity in LaFeAsO$_{1-x}$F$_x$ (Ref. \onlinecite{Kamihara08}), high transition temperatures $T_c$ up to 56~K were reported for several Fe-based superconductors with La substituted by other lanthanoids ({\it Ln}) including e.g., Ce, Pr, Nd, Sm, and Gd.\cite{XHChen08,GFChen08,Ren08,Ren08a,Cheng08}
Meanwhile, the family of Fe-based superconductors range from {\it Ln}FeAsO$_{1-x}$F$_x$ (the so called 1111 family) over {\it Ae}Fe$_2$As$_2$ (122, {\it Ae}~=~alkaline earth metal)\cite{Rotter08} to the more simple LiFeAs (111)\cite{Wang08} and Fe{\it Ch} (11, {\it Ch}~=~chalcogenide).\cite{Hsu08} The Fe{\it Ch} system is especially similar to the FeAs-based superconductors, reflecting the ionic nature of the As and chalcogen atoms in these compounds.\cite{Subedi08} Recently, two  even more complicated families were discovered: the (Fe$_2$As$_2$)({\it Ae}$_4${\it M}$_2$O$_6$) (22426, {\it M}~=~transition metal) and the (Fe$_2$As$_2$)({\it Ae}$_3${\it M}$_2$O$_5$) (22325) systems.\cite{Ogino09,Zhu09}
If the parent compound is not already superconducting, superconductivity can be induced by charge carrier doping into either the Fe layers or the spacer layers as well as by applying external or internal pressure.\cite{Luetkens09,Drew09,Mizugushi08,KhasanovFeSeTe}

Fe-based superconductors share some common properties with high-$T_c$ cuprates such as a layered crystal structure, the presence of competing orders, a low carrier density, a small coherence length, and an unconventional pairing mechanism.
On the other hand, there are some differences: The Fe-based superconductors have metallic parent compounds, the anisotropy is in general lower compared to that of the cuprates, and the order parameter symmetry is claimed to be $\pm s$-wave with Fermi-surface nesting playing a major role.\cite{Karpinski09,Mazin09,Amato09,Graser09}
So, the fundamental question arises whether the mechanisms leading to superconductivity in both families of high-temperature superconductors (HTS) share a common origin.

Among the Fe-based superconductors the "11" system has attracted a lot of attention. The transition temperature $T_c$ of FeSe$_{1-x}$ reaches values up to $\approx 37$~K by applying hydrostatic pressure\cite{Hsu08,Margadonna09} and $\approx 14$~K by partially substituting Se by the isovalent Te or S.\cite{sales09} In FeSe$_{x}$Te$_{1-x}$ the antiferromagnetic order of FeTe is gradually suppressed by increasing $x$, and superconductivity emerges with a maximal $T_c$ at $x\simeq0.5$.\cite{KhasanovFeSeTe}
Additionally, the "11" system has the simplest crystallographic structure among the Fe-based superconductors consisting of layers with a Fe square planar sheet tetrahedrally coordinated by {\it Ch}.\cite{Hsu08}
This and the similarity of the Fermi surface to the one of the FeAs based superconductors\cite{Subedi08} make the ''11'' system an ideal candidate to study the interplay of structure, magnetism, and superconductivity in Fe-based superconductors.
In this paper we report on the structural and anisotropic superconducting properties of single crystals with the nominal composition of FeSe$_{0.5}$Te$_{0.5}$ that were studied by neutron powder diffraction, SQUID and torque magnetometry as well as muon spin rotation ($\mu$SR). A part of the present results are in agreement with the findings of a recent $\mu$SR study performed on a polycrystalline sample of FeSe$_{0.5}$Te$_{0.5}$.\cite{Biswas10}

\section{Experimental details}
\subsection{Single crystal growth}
Single crystals with the nominal composition of FeSe$_{0.5}$Te$_{0.5}$ were grown by the Bridgman method, similar to that reported by Sales {\it et al}.\cite{sales09}
Appropriate amounts of Fe, Se, and Te powders with a minimum purity of 99.99 \% were mixed together, pressed into a rod (diameter 7 mm), and than evacuated and sealed in a double-wall quartz ampoule for air protection.
The ampoule was placed into a vertical furnace with a temperature gradient and annealed at 1200~$^\circ$C for 4~h. Afterwards the samples were cooled down with a rate of 4~$^\circ$C/h to 750~$^\circ$C, followed by a quick cooling (50~$^\circ$C/h) to room temperature.
The so-obtained crystals were easily cleaved from the as-grown crystal along the $ab$-plane (cleaving facet).

\begin{figure}[t!]
\centering
\vspace{-0cm}
\includegraphics[width=1\linewidth]{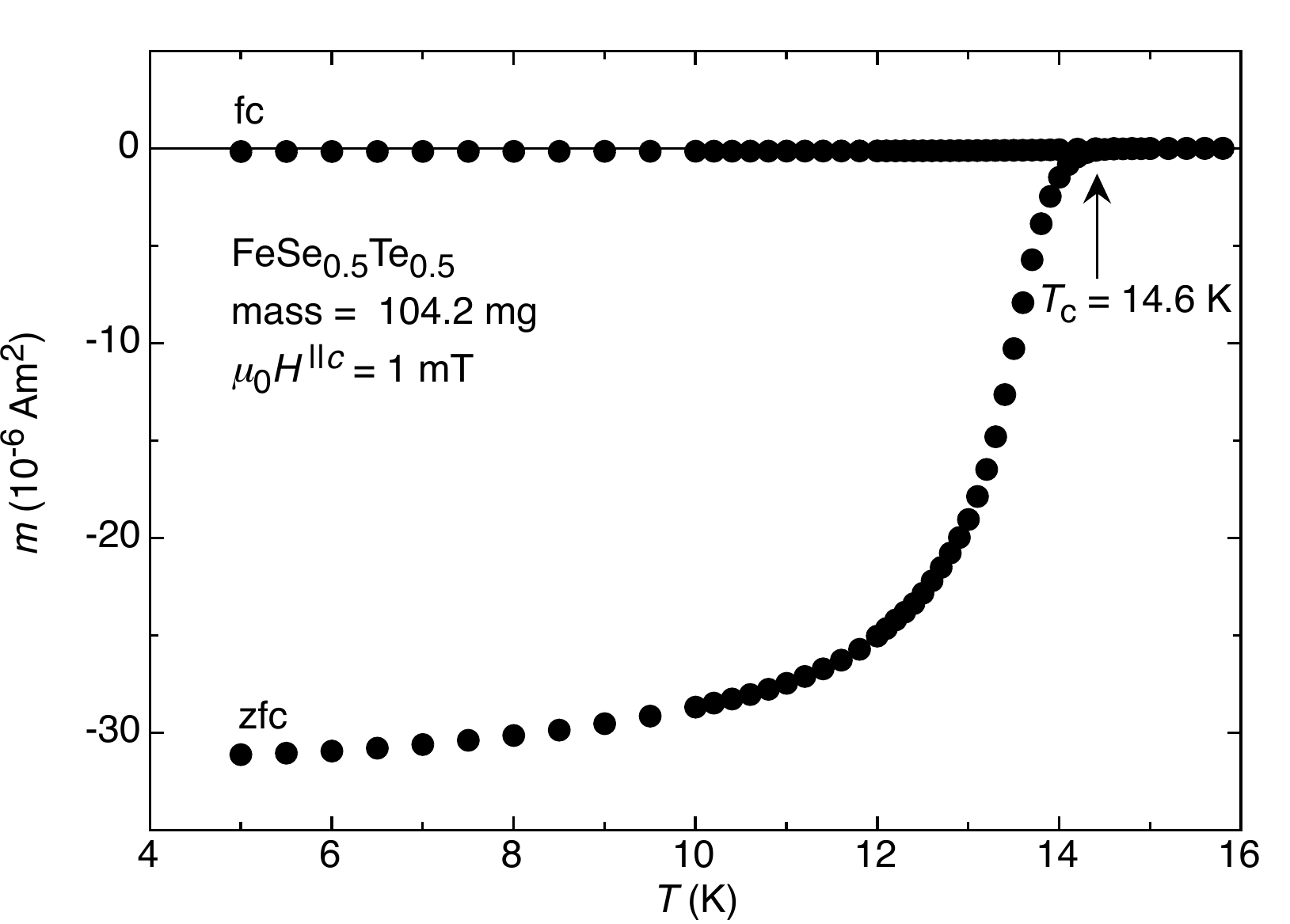}
\caption{Magnetic moment $m$ as a function of temperature $T$ in a magnetic field of 1~mT applied parallel to the $c$-axis of single-crystal FeSe$_{0.5}$Te$_{0.5}$, recorded in the Meissner state in zero field cooled (zfc) and in field cooled (fc) mode. The onset transition temperature $T_c$ $\simeq14.6$~K (vertical arrow) is characteristic for optimal doping $x\simeq 0.5$ of FeSe$_{x}$Te$_{1-x}$.}
\label{fig_Tc}
\end{figure}

\begin{figure}[t!]
\centering
\vspace{-0cm}
\includegraphics[width=1.1\linewidth]{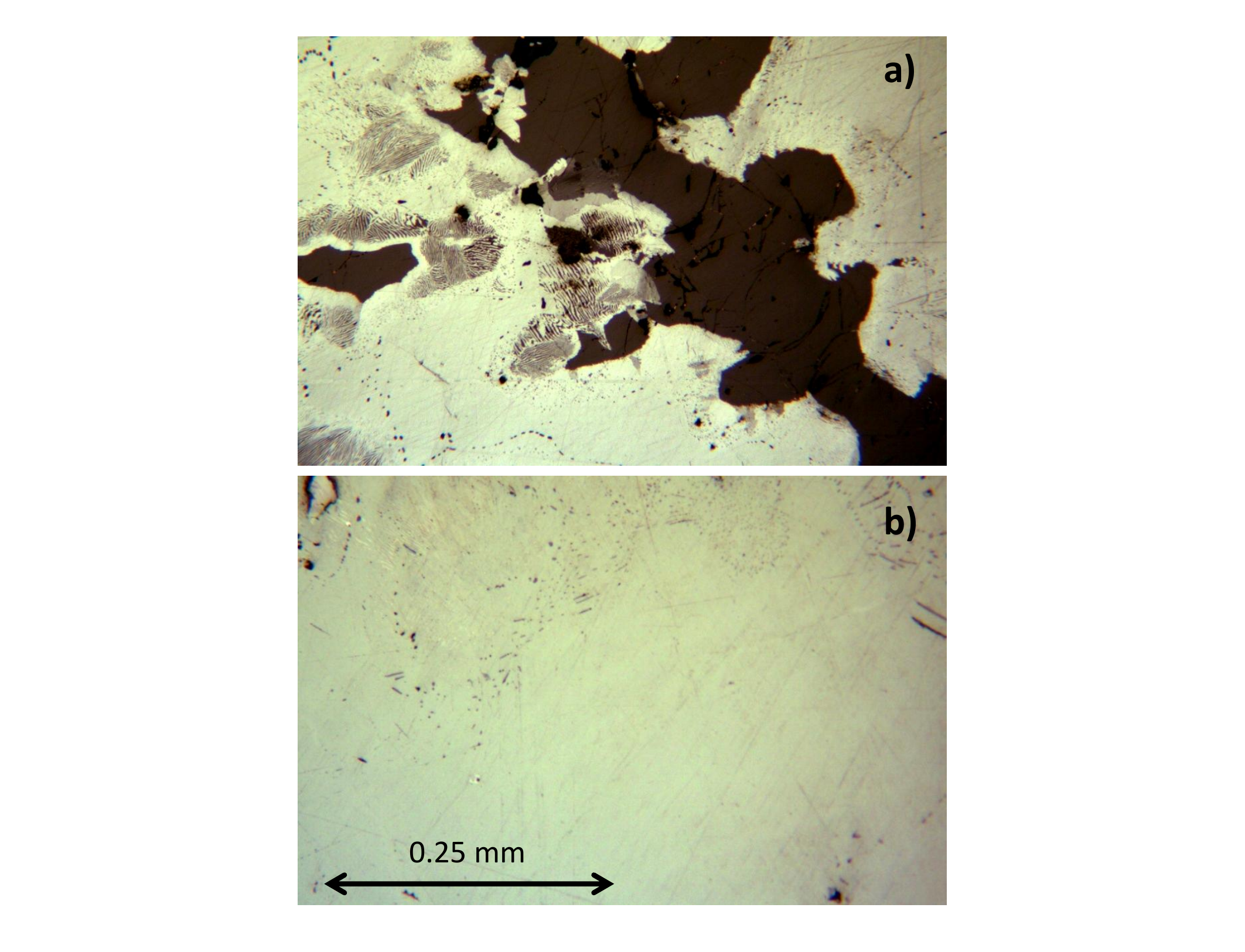}
\caption{(color online) Polarized light microscopic photographs of polished surfaces of the FeSe$_{0.5}$Te$_{0.5}$ crystal. a) Microphotography of the crystal surface cut perpendicular to the cleaving facet. Domains with different crystallographic orientations and/or different phases are visible. b) Micrography of the resulting polished cleaving facet.}
\label{fig_microscopy}
\end{figure}

Figure \ref{fig_Tc} presents a low-field measurement of the magnetic moment $m$ in a magnetic field of $\mu_0H=1$~mT applied along the $c$-axis performed in zero field cooled (zfc) and field cooled (fc) mode. The sample exhibits a clear transition to the superconducting state with an onset transition temperature of $T_{\rm c}\simeq14.6$~K. The signal magnitude obtained in the zfc mode reflects a full diamagnetic response of the sample. The low value of the fc signal indicates strong pinning.\\
The surface of the as-grown crystal was polished, and the surface morphology was examined in a polarized light microscope. Figure \ref{fig_microscopy}a shows a microphotography of the crystal surface cut perpendicular to the cleaving facet. Distinct domains of different crystallographic orientations and/or different phases are observed.
Figure \ref{fig_microscopy}b shows the polished cleaving facet.
No orientation misfit is observed here.
In conclusion, the main phase in the material is textured with the {\it c}-axis perpendicular to the cleaving facet, whereas the {\it a}- and {\it b}-axes are oriented within domains of irregular shape.\\

\subsection{Crystal structure}
\begin{figure}[t!]
\centering
\vspace{-0cm}
\includegraphics[width=1\linewidth]{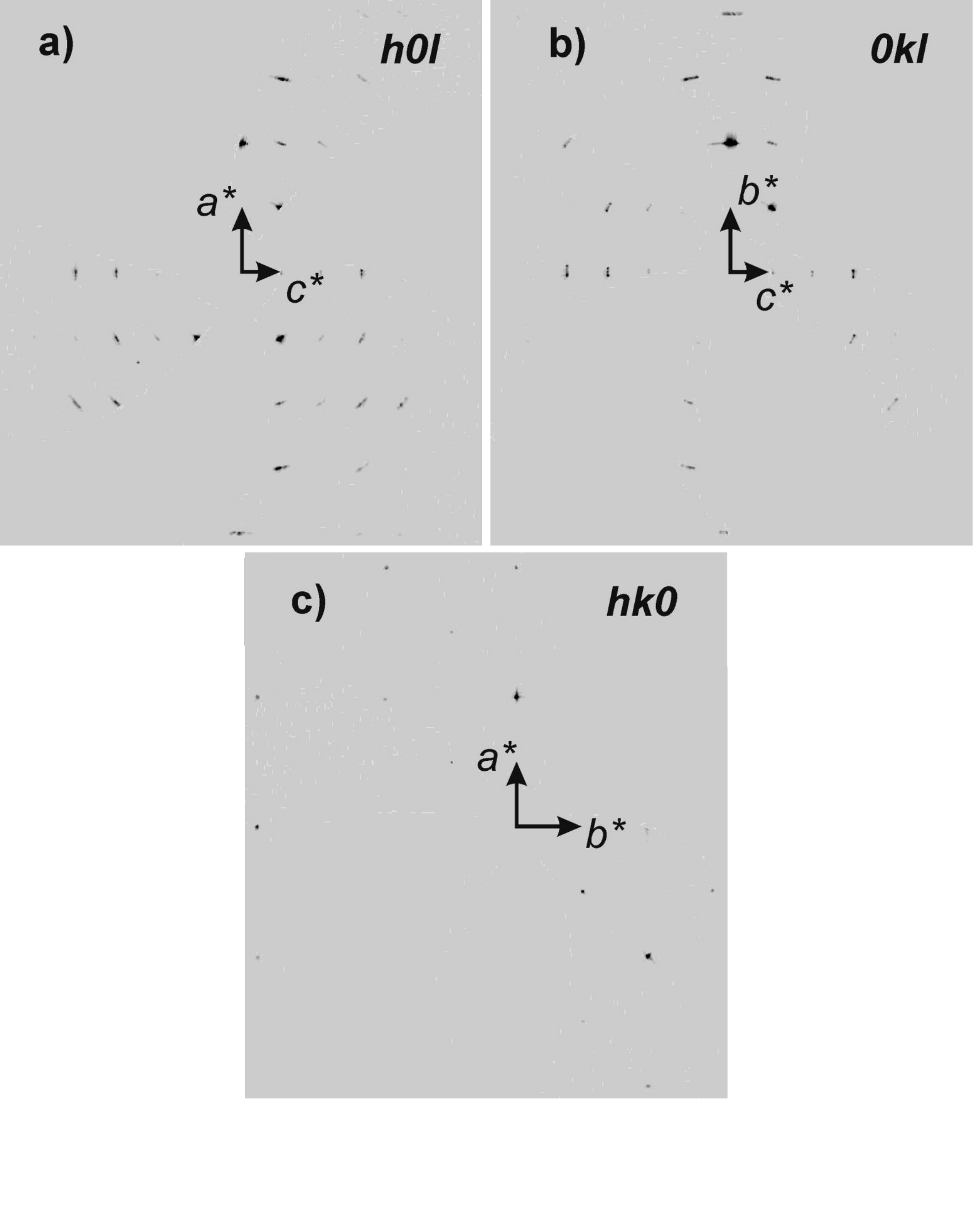}
\caption{(color online) The reciprocal space sections of the FeSe$_{0.5}$Te$_{0.5}$  crystal: a) $h0l$ reciprocal layer; b) $0kl$ reciprocal layer; c) $hk0$ reciprocal layer.}
\label{fig_XRD}
\end{figure}

The crystal structure and the phase purity were checked using a single crystal X-ray diffractometer equipped with a charged-coupled device (CCD) detector (Xcalibur PX, \textit{Oxford Diffraction}, sample-detector distance $60$~mm).
Crystallites with approximate dimensions of $1\times1\times0.2$~mm$^3$ were cleaved from the as-grown crystal for the single crystal X-ray diffraction studies.
The single crystal diffractographs are shown in Fig. \ref{fig_XRD}.
Two distinct crystallographic phases were identified.
The major phase of the crystal exhibits a tetragonal lattice (space group: $P4/nmm$, lattice parameters: $a=3.7980(2)$~\AA, $c=6.038(1)$~\AA).
The reconstruction of the reciprocal space sections of the studied plate-like crystals lead to pronounced mosaic spreads with an average mosaicity of the order of about $4^{\circ}$.
A small part of the studied crystals with polygonal structure exhibits a hexagonal lattice structure, which is associated with an impurity phase.

\begin{figure}[t!]
\centering
\vspace{-0cm}
\includegraphics[width=1\linewidth]{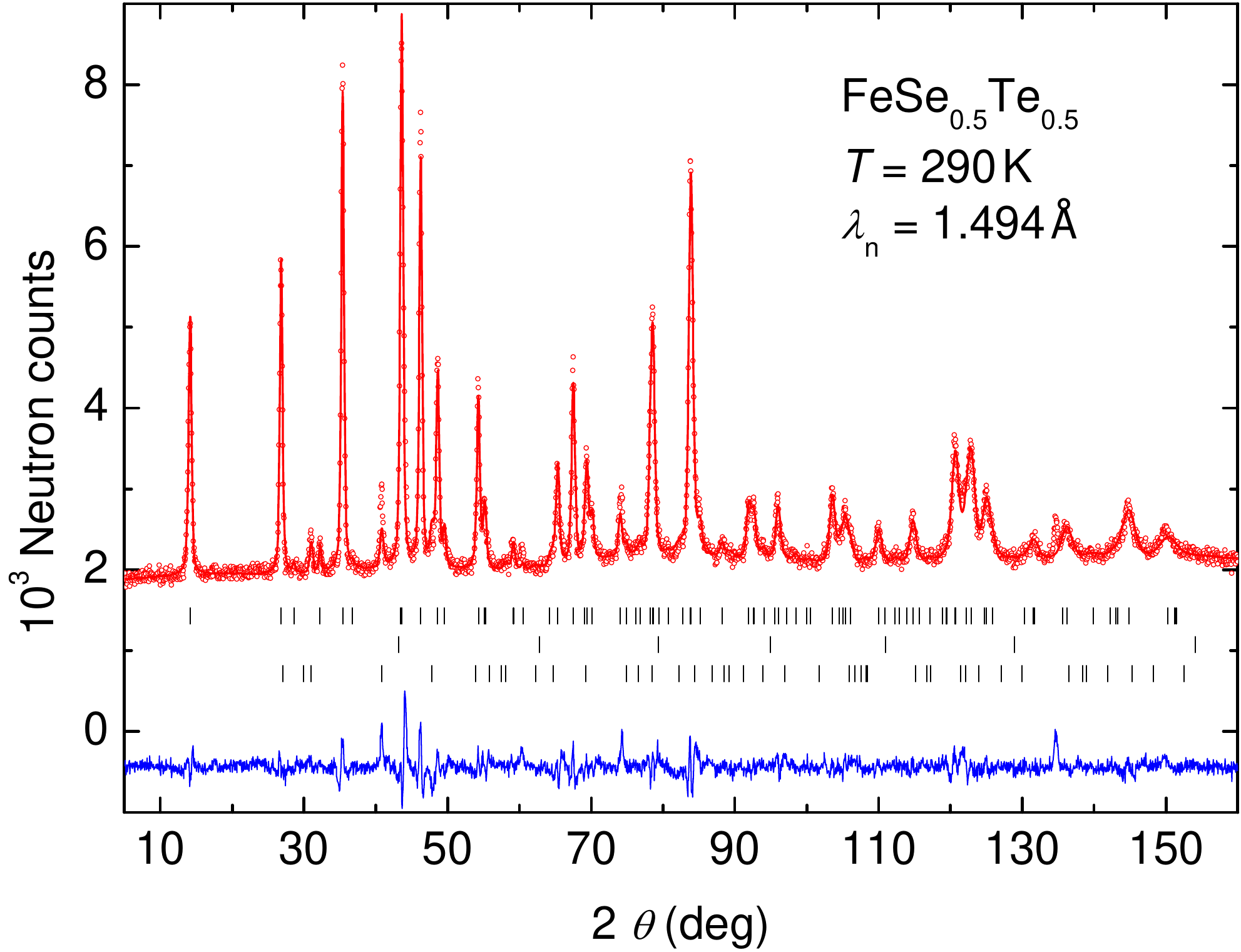}
\caption{(color online) Rietveld refinement pattern (red) and difference plot (blue) of the neutron diffraction data for the crystal with the nominal composition of FeSe$_{0.5}$Te$_{0.5}$. The rows of ticks show the Bragg peak positions for the main phase and two impurity phases. The refined stoichiometry of the main tetragonal phase is Fe$_{1.045}$Se$_{0.406}$Te$_{0.594}$ (see text for details).}
\label{fig_NPD}
\end{figure}

Detailed crystal structure investigations were completed by means of neutron powder diffraction (NPD) at the neutron spallation source SINQ at the Paul Scherrer Institute (PSI, Switzerland) using the high resolution powder diffractometer for thermal neutrons HRPT \cite{HRPT} (neutron wavelength $\lambda_{\rm n}=1.494$~\AA).
For these experiments a part of the crystal with the nominal composition of FeSe$_{0.5}$Te$_{0.5}$ was cleaved, powderized, and loaded into the sample holder in a He-glove box to protect the powder from oxidation.
Room temperature NPD experiments revealed that the sample consists mainly of the tetragonal phase (space group $P4/nmm$) of the PbO type which becomes orthorhomic and superconducting at low temperatures. The results of the Rietveld refinement of the NPD spectra performed with the program FULLPROF\cite{FULLPROF} are shown in Fig. \ref{fig_NPD}.
For the refinement it was assumed that all Fe sites are occupied. Additionally, a preferred orientation was assumed as small plate-like grains are created during the powderization process.
The refined stoichiometry is Fe$_{1.045}$Se$_{0.406(16)}$Te$_{0.594(16)}$ ($a=3.8028(1)$~\AA, $c=6.0524(3)$~\AA).
Note that these values were obtained by assuming a texture in the powder sample.
As impurity phases hexagonal Fe$_7$Se$_8$ (space group $P6_{3}/mmc$, $5.35(40)$\% volume fraction) and elemental Fe ($\leq 1$\%) were identified.\\
It was shown that in the $\beta$-phase additional excess Fe occupies interstitial lattice sites.\cite{Li09,Bao09} However, introduction of interstitial Fe atoms in the refinement of the data did not improve the fit.
This suggests the presence of only a very small amount of such defects, in agreement with the model that in isostructural FeSe$_{1-x}$\cite{Pomjakushina09} no interstitial Fe is present. This is in contrast to FeTe where interstitial Fe atoms were detected.\cite{Li09,Bao09}  \\
\begin{figure}[t!]
\centering
\vspace{-0cm}
\includegraphics[width=1\linewidth]{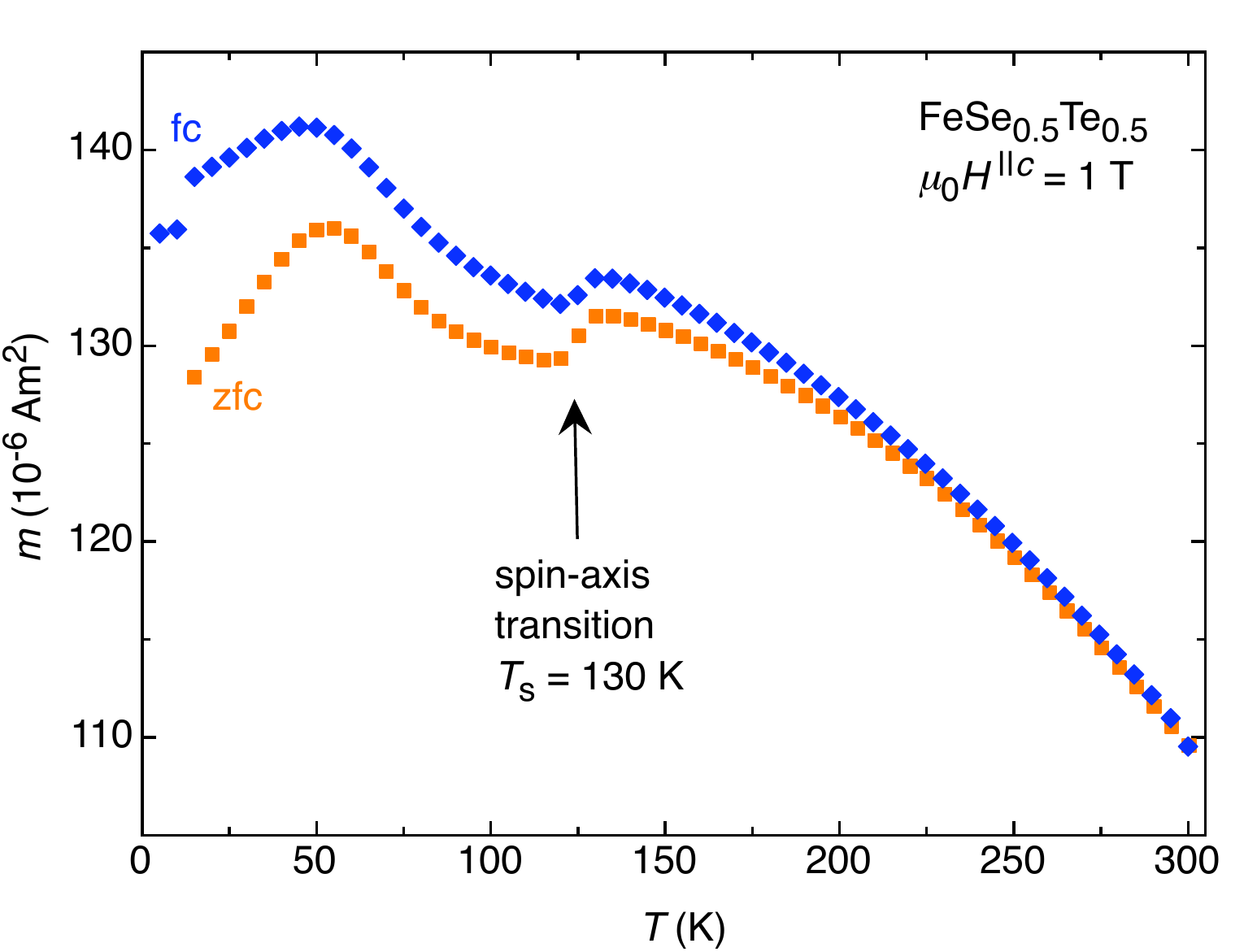}
\caption{(color online) Temperature dependence of the magnetic moment measured in zero field cooling (zfc) and field cooling (fc) mode in a magnetic field of 1~T applied parallel to the $c$-axis of the crystal with nominal composition FeSe$_{0.5}$Te$_{0.5}$. }
\label{fig_magn_Fe7Se8}
\end{figure}
The existence of an impurity phase of Fe$_{7}$Se$_{8}$ in the studied crystal was confirmed by magnetization measurements. Figure \ref{fig_magn_Fe7Se8} shows the temperature dependence of the magnetic moment recorded for a FeSe$_{0.5}$Te$_{0.5}$ crystal (mass $\sim 200$~mg) in a magnetic field of 1~T, applied parallel to the $c$-axis of the crystal.
Fe$_{7}$Se$_{8}$ is known to undergo a spin-axis transition at 130~K leading to a reduction of magnetization for $H$ parallel to the $c$-axis,\cite{Fe7Se8} as observed in the studied sample (Fig. \ref{fig_magn_Fe7Se8}).

\section{Magnetic properties}
\subsection{Magnetization measurements}

The magnetic properties of the crystals were investigated by a commercial \textit{Quantum Design} $7$~T Magnetic Property Measurement System (MPMS) XL SQUID Magnetometer at temperatures ranging from $2$~K to $300$~K and in magnetic fields from $0$~T to $7$~T using the Reciprocating Sample Option (RSO). Magnetic torque measurements were performed with a commercial \textit{Quantum Design} $9$~T Physical Property Measurement System (PPMS) equipped with a magnetic torque option.

\begin{figure}[t!]
\centering
\vspace{-0cm}
\includegraphics[width=1\linewidth]{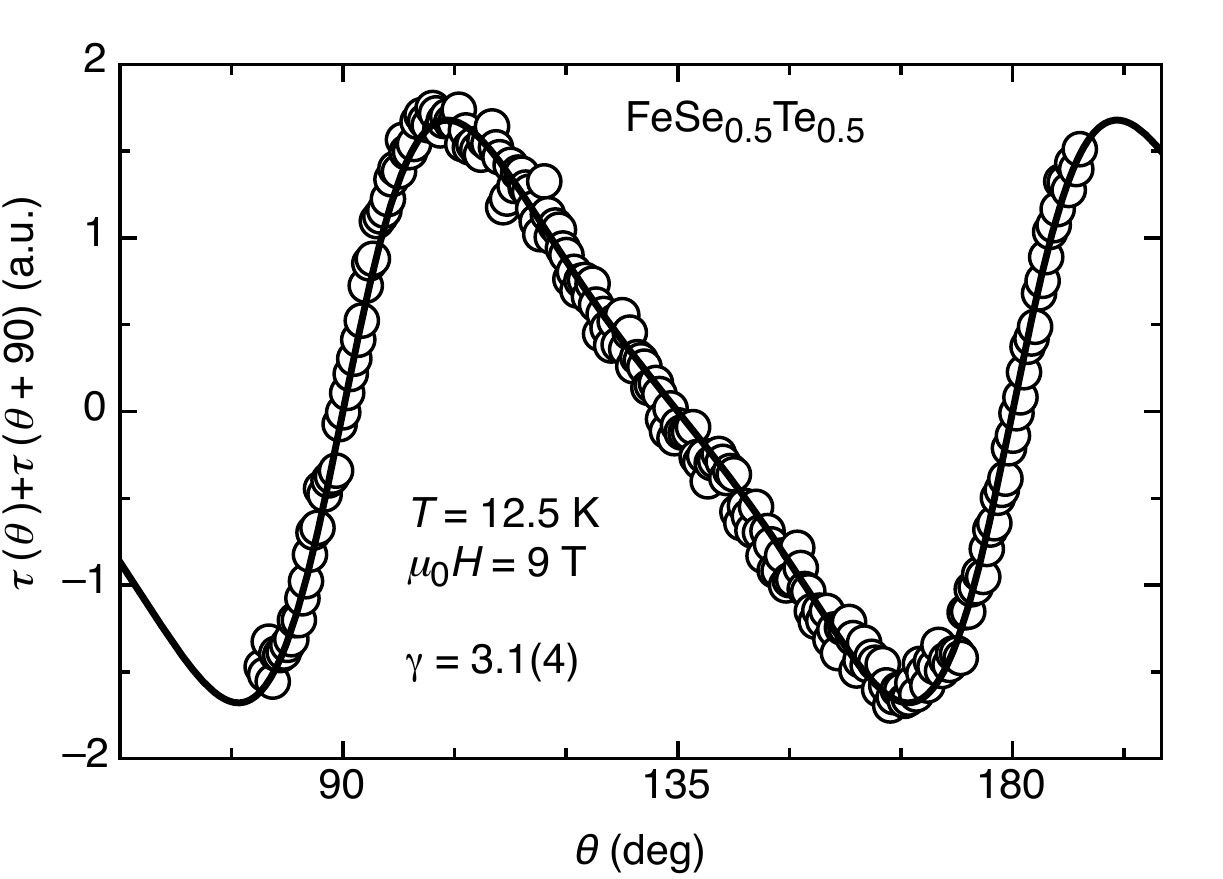}
\caption{Symmetrized torque $\tau_{symm}$ for the studied crystal of FeSe$_{0.5}$Te$_{0.5}$ in the superconducting state as a function of the angle $\theta$. The torque data are well described by Eq. (\ref{eq_torque}), yielding an anisotropy parameter $\gamma=3.1(4)$ close to $T_c$.}
\label{fig_torque}
\end{figure}

The magnetization of FeSe$_{0.5}$Te$_{0.5}$ was measured on a crystal with a mass of the order of 200~mg. The Meissner fraction derived from the magnetic moment in the fc mode as compared to the one from zfc is estimated to be $\sim1$~\% in $1$~mT (Fig. \ref{fig_Tc}). This indicates strong vortex pinning in agreement with the weakly field-dependent and pronounced critical current denstity and with the significant irreversibility in the magnetic torque experiments already present slightly below $T_c$ (as discussed later, Fig. \ref{fig_torque}). Using Bean's model\cite{Bean62,Bean64} magnetization hysteresis loop measurements allow to estimate the superconducting critical current of the order of $10^7$~A/m$^2$. The presence of impurity phases is lowering the transport curret density as phase separation boundaries prevent to develop a global circulating current. This leads to a relatively low value of the estimated critical current density as compared to those achieved for monocrystalline iron-pnictides.\cite{Karpinski09}

\begin{figure}[t!]
\centering
\vspace{-0cm}
\includegraphics[width=1\linewidth]{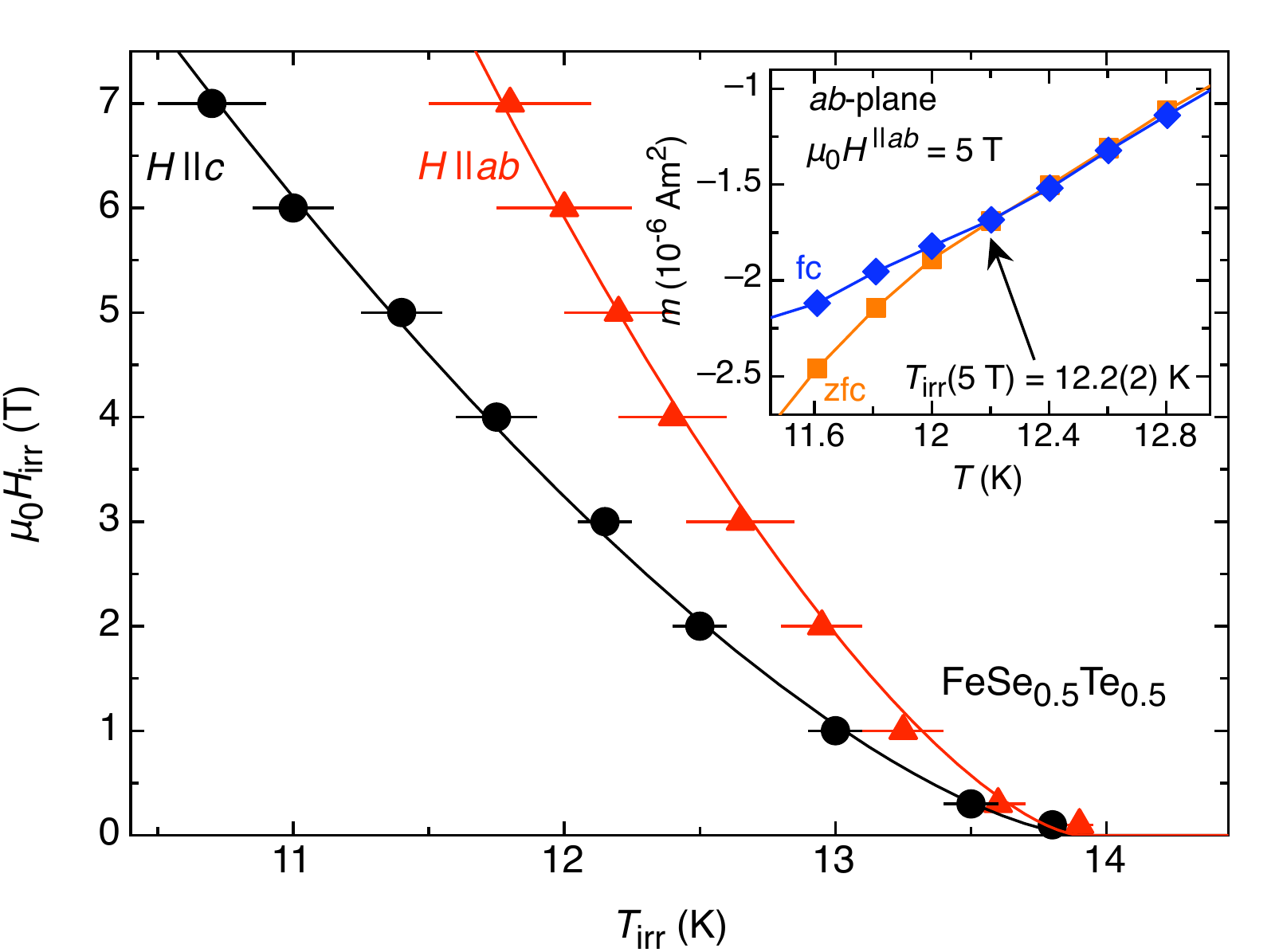}
\caption{(color online) Irreversibility line $H_{\rm irr}$($T$) derived from SQUID measurements for two field configurations, $H$ parallel to the $c$-axis and $H$ parallel to the $ab$-plane of the FeSe$_{0.5}$Te$_{0.5}$ crystal. The solid black and red lines correspond to fits using the power-law $(1-T/T_c)^n$ with an exponent $n\simeq 1.5$. The inset illustrates how $H_{\rm irr}$ was determined. The lines are guides to the eyes.}
\label{fig_irreversibility}
\end{figure}

\begin{figure}[t!]
\centering
\vspace{-0cm}
\includegraphics[width=1\linewidth]{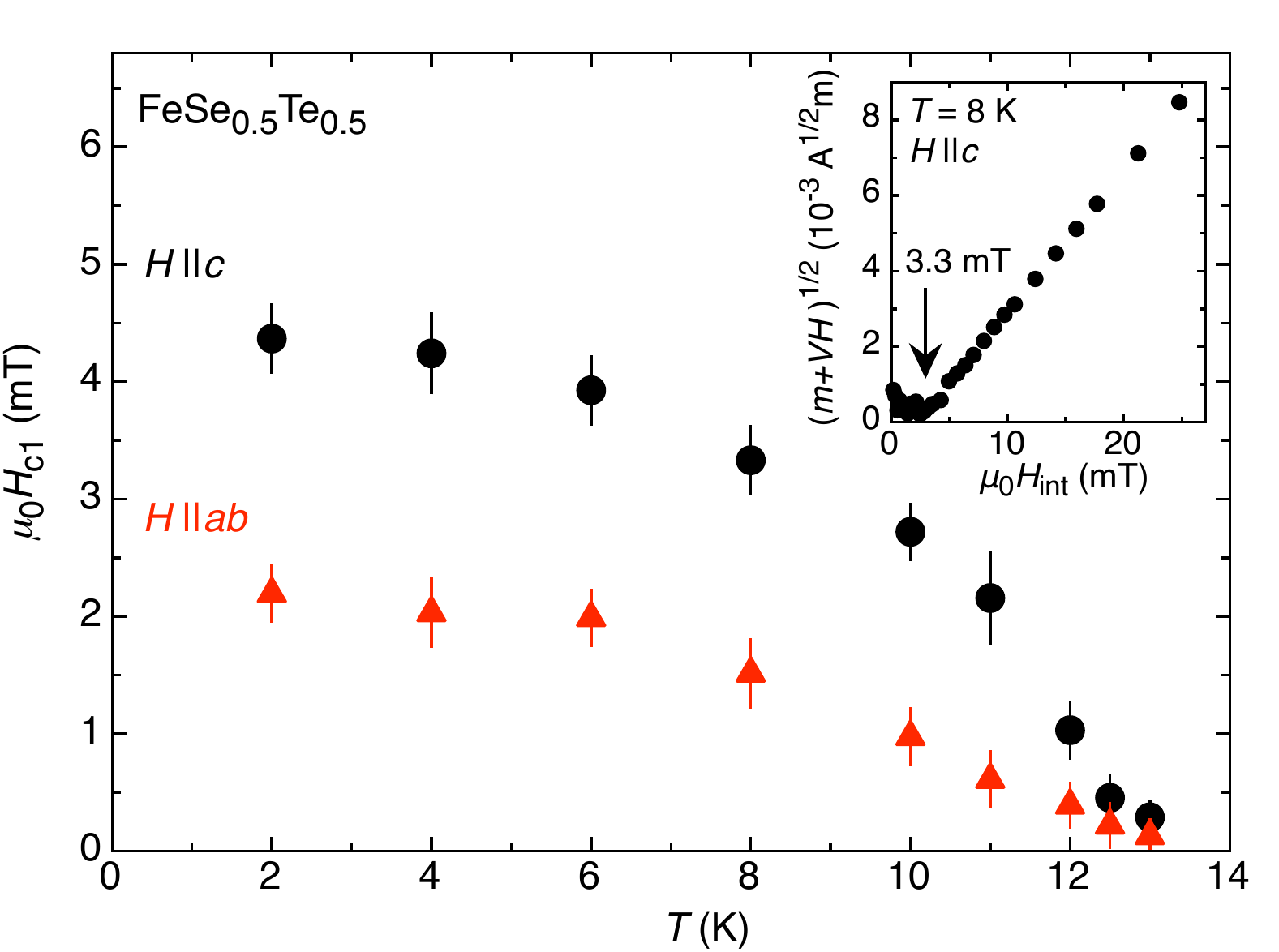}
\caption{(color online) $H_{\rm c1}$ as a function of temperature for both orientations $H$ parallel to the $c$-axis and $H$ parallel to the $ab$-plane for single-crystal FeSe$_{0.5}$Te$_{0.5}$. The inset illustrates the deviation from the linear $B^{1/2}$($H$) dependence plotted as $B^{1/2}$ vs. $H_{\rm int}$.}
\label{fig_Hc1_T}
\end{figure}

From temperature-dependent magnetization measurements at various magnetic fields the irreversibility line $H_{\rm irr}$($T$) was deduced by following the temperatures for which the zfc and fc branches of the magnetic moment merge for different fields. The results are presented in Fig. \ref{fig_irreversibility}, where the inset to the figure illustrates the derivation of $H_{\rm irr}$ in a magnetic field of $5$~T parallel to the $ab$-plane. The data were analyzed using the power-law  $(1-T/T_c)^{n}$ with $n\simeq 1.5$, typical for cuprate HTS.\cite{Yeshurun88} The irreversibility line $H_{\rm irr}$ is located at relatively high magnetic fields. Interestingly, $H_{\rm irr}$ is for $H$ parallel to the $ab$-plane almost overlapping with the values of the upper critical field $H_{\rm c2}^{\parallel c}$ reported by Fang {\it et al}.\cite{fang09}\\
The temperature dependence of the lower critical field $H_{\rm c1}$ was studied by following the field $H_{\rm c1}^*$, where the first vortices start to penetrate the sample at its surface, which is directly related to $H_{\rm c1}$.
The field dependence of the magnetization was measured at different temperatures for the magnetic field parallel to the $ab$-plane and parallel to the $c$-axis of the sample.
For a given shape of the investigated crystal the demagnetizing factors $D$ were calculated for the magnetic field applied along all of the crystallographic axis. 
The deviation of the magnetic induction $B$ as a function of the internal magnetic field $H_{\rm int} = H_{\rm ext} - DM$ ($H_{\rm ext}$ denotes the external magnetic field) is presented in Fig. \ref{fig_Hc1_T}.
The lower critical fields for $H$ parallel to the $ab$-plane and parallel to the $c$-axis presented in Fig. \ref{fig_Hc1_T} were determined as the field where the magnetization deviates from the linear behavior. From these data the zero temperature values were found to be $\mu_0 H_{\rm c1}^{\parallel ab}(0)=2.0(2)$~mT and $\mu_0 H_{\rm c1}^{\parallel c}(0)=4.5(3)$~mT. In order to extract the values of the magnetic penetration depth from the measured values of $H_{\rm c1}$ the following basic relations were applied:\cite{Tinkham}
\begin{eqnarray}
H_{{\rm c}1}^{||c}&=&\frac{\Phi_0}{8 \pi \mu_{0} \lambda_{ab}^2}\left[2\ln\left(\frac{\lambda_{ab}}{\xi_{ab}}\right)+1\right],\label{eq_Hc1a}\\
H_{{\rm c}1}^{||ab}&=&\frac{\Phi_0}{8 \pi \mu_{0} \lambda_{ab}\lambda_c}\left[\ln\left(\frac{\lambda_{ab}\lambda_c}{\xi_{ab}\xi_c}\right)+1\right].
\label{eq_Hc1b}
\end{eqnarray}
Here, $\lambda_{ab}$ and $\lambda_c$ are the magnetic penetration depths parallel to the $ab$-plane and to the $c$-axis, respectively, $\xi_{ab}$ and $\xi_c$ the corresponding coherence lengths, $\Phi_0$ is the elementary flux quantum, and $\mu_0$ the magnetic constant.
The values of $\xi_{ab}$ and $\xi_c$ were derived from $H_{\rm c2}^{\parallel ab}$ and $H_{\rm c2}^{\parallel c}$ measurements.\cite{fang09}
The following zero temperature values of magnetic penetration depths were obtained: $\lambda_{ab}(0)\simeq460(100)$~nm and $\lambda_{c}(0)\simeq1100(300)$~nm. These values are in good agreement with the values determined by $\mu$SR discussed below (see Table \ref{table_twogap}).

In order to quantify the anisotropy of superconducting state parameters, magnetic torque studies were performed close to $T_c$, where irreversibility effects are small.
The measurements on small crystals ($\sim 1 \times 1 \times 0.2$~mm$^3$) revealed a major superconducting response, in agreement with the NPD results discussed above.
Unfortunately, due to the small amplitude of the superconducting torque signal in the mixed state close to $T_c$, a relatively strong background component of magnetic origin is contributing significantely to the torque signal.
The magnetic background signal in the superconducting state is confirmed by following the torque to temperatures above $T_{\rm c}$.
In order to exclude artefacts in the subsequent analysis, all background components within the superconducting state were subtracted from the torque prior to the analysis (see below).
To minimize the influence of pinning the mean reversible torque $\tau_{\rm rev}=[\tau(\theta^+)+\tau(\theta^-)]/2$ was derived from measurements with clockwise and counterclockwise rotating the magnetic field around the sample.
The superconducting anisotropy parameter $\gamma=\lambda_{c}/\lambda_{ab}$ may be extracted from the measured torque $\tau(\theta)$ using the relation:\cite{Kogan81,Kogan88}
\begin{eqnarray}
\nonumber\tau(\theta)= -\frac{V\Phi_{0}H}{16 \pi \lambda^{2}_{ab}}\left(1-\frac{1}{\gamma^{2}} \right) \frac{\sin(2\theta)}{\epsilon(\theta)}\\ \times \ln \left(\frac{\eta H^{\parallel c}_{\rm c2}}{\epsilon(\theta)H} \right)
 + A_{\tau}\sin(2\theta),
 \label{eq_torque}
\end{eqnarray}
where $V$ is the volume of the crystal, $\lambda_{ab}$ is the in-plane component of the magnetic penetration depth, $H_{\rm c2}^{\parallel c}$ is the upper critical field along the $c$-axis of the crystal, $\eta$ denotes a numerical parameter of the order of unity depending on the structure of the flux-line lattice, $A_{\tau}$ is the amplitude of the background torque, and $\epsilon(\theta)=[\cos^2(\theta)+\gamma^{-2}\sin^2(\theta)]^{1/2}$.
Since Eq. (\ref{eq_torque}) contains multiple correlated parameters, making a simultaneous fit of all quantities difficult, all $H_{\rm c2}^{\parallel c}$ values were fixed to those reported in Ref. \onlinecite{fang09} during the fitting procedure by neglecting any influence of the parameter $\eta$.
Because the magnetic background contributions tend to influence and alter the fitting parameter $H_{\rm c2}^{\parallel c}$ strongly,\cite{Weyeneth09,Weyeneth09a} the data were fitted by Eq. (\ref{eq_torque}) using the symmetrized expression for the torque $\tau_{\rm symm}(\theta)=\tau(\theta)+\tau(\theta+90^{\circ})$.\cite{Balicas09} The result of this analysis is depicted in Fig. \ref{fig_torque}, yielding an anisotropy parameter $\gamma=3.1(4)$ in the vicinity of $T_{c}$.\\

\subsection{Muon spin rotation}
Muon spin rotation ($\mu$SR) is a direct and bulk sensitive probe to investigate local magnetic fields in magnetic solids.\cite{Zimmermann95}
Nearly 100\% spin-polarized positive muons $\mu^+$ are implanted into the sample and stop at interstitial lattice sites, where the muon spins precess around the local magnetic field $B$ with the Larmor frequency $\omega_{\rm L}=\gamma_\mu B$ ($\gamma_\mu/2\pi=135.5$~MHz/T is the muon gyromagnetic ratio). At the stopping site the muon acts as a magnetic micro probe and measures the internal field distribution. Within the muon's life time of $\tau=2.2$~$\mu$s it decays into two neutrinos and a positron, which is emitted predominantly along the muon spin polarization at the moment of decay. The direction of the emitted decay positron and the time between the muon implantation and its decay is measured for typically $10^6$ muons. This way the time evolution of the muon spin polarization $P(t)$ is obtained. Zero-field (ZF) $\mu$SR experiments probe the magnetic state of a material as the muon spins precess only around the internal field without applying an external magnetic field. In transverse field (TF) $\mu$SR experiments the local magnetic field at the muon site in the sample is probed in the presence of an external magnetic field perpendicular to the initial muon spin polarization. TF $\mu$SR is a very powerful tool to investigate the local magnetic field distribution in the vortex state of type II superconductors. A comprehensive review of the application of $\mu$SR to the study of superconductors can be found in Ref. \onlinecite{Zimmermann95}.

The $\mu$SR experiments were carried out at the $\pi$M3 beam line at the Swiss Muon Source (S$\mu$S) at PSI. ZF and TF $\mu$SR experiments were performed in a temperature range from $1.5$~K to $20$~K. The TF experiments were carried out in two sets of measurements when the external field $\mu_0H=11.8$~mT was applied either parallel to the crystallographic $c$-axis or parallel to the $ab$-plane.

\begin{figure}[t!]
\centering
\vspace{-0.4cm}
\includegraphics[width=1\linewidth]{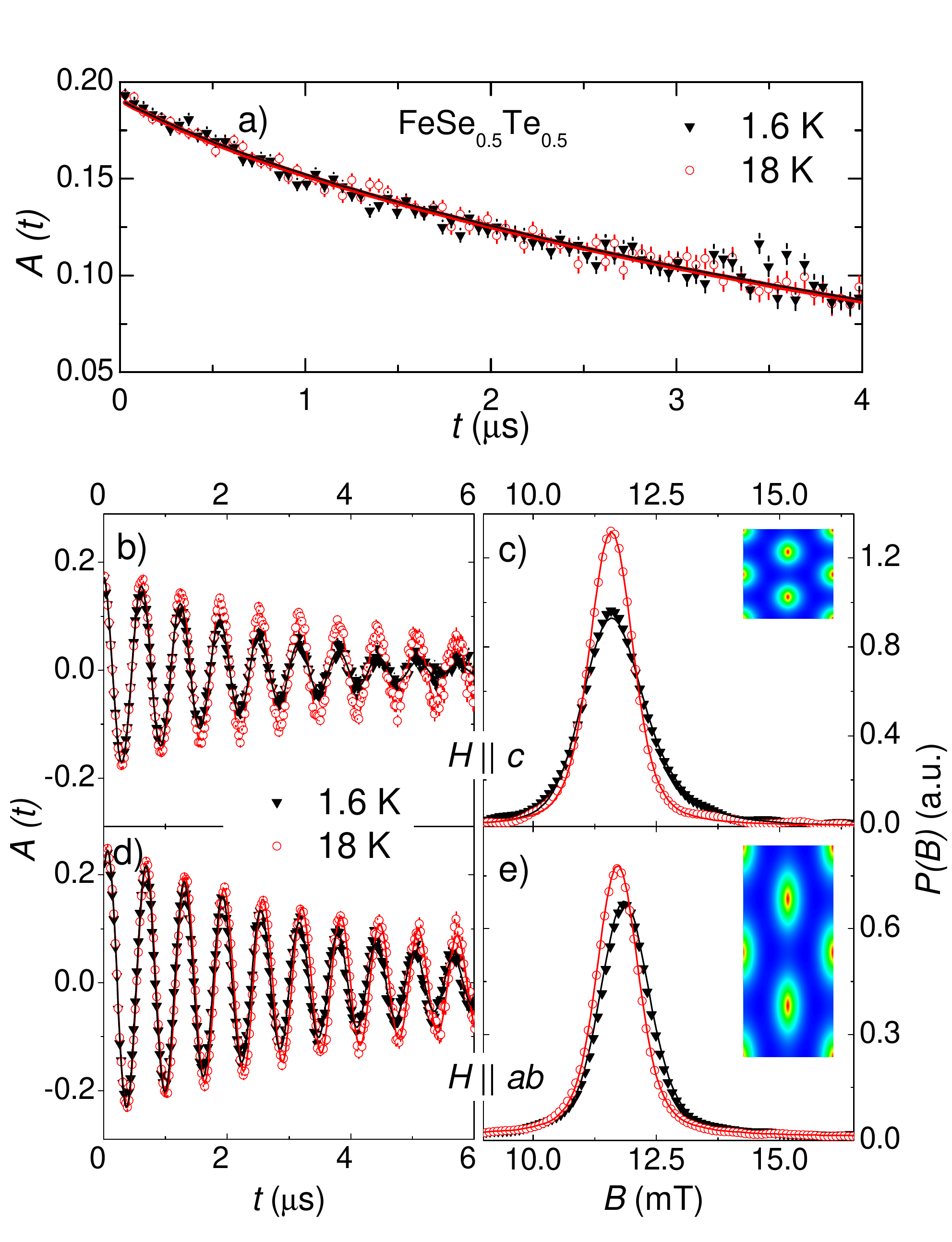}
\caption{(color online) a) ZF $\mu$SR time spectra for FeSe$_{0.5}$Te$_{0.5}$ recorded at $1.6$~K and above $T_c$. The solid lines represent fits using Eq. (\ref{eq_ZF}). b), d) TF $\mu$SR time spectra for $H$ parallel to the $c$-axis and $H$ parallel to the $ab$-plane, taken at $1.6$~K and above $T_c$. c), e) The corresponding magnetic field distributions $P(B)$. The solid lines represent fits using Eq. (\ref{eq_TF}). The insets show the counter plots of the local field variation at 1.6~K, c) $\lambda_{a}=\lambda_b$; e) $\lambda_{c}=2.7\lambda_{ab}$.}
\label{fig_ZF_raw}
\end{figure}

The ZF $\mu$SR spectra obtained at $1.6$~K and above $T_c$ show no difference (Fig. \ref{fig_ZF_raw}a). This indicates that the magnetic state of FeSe$_{0.5}$Te$_{0.5}$ below and above $T_c$ is the same. The solid lines in Fig. \ref{fig_ZF_raw}a correspond to a fit using an exponential decay of the initial muon spin polarization:
\begin{equation}
A^{\rm ZF}(t)=A_{\rm SC} \cdot e^{-\Lambda t}+A_{\rm bg} \cdot e^{-\Lambda_{\rm bg} t}.
\label{eq_ZF}
\end{equation}
Here $A_{\rm SC}$ is the asymmetry of the superconducting phase and $\Lambda$ is the corresponding depolarization rate.
The temperature independent background signal $A_{\rm bg}$, arising from the Fe$_7$Se$_8$ impurity phase was fixed to 6\% of the total asymmetry during the fit, corresponding to the results of the NPD refinement.
The exponential character of the muon spin depolarization is typical for diluted and randomly distributed magnetic moments, that are static on the muon time scale as shown in Ref. \onlinecite{khasanovFeSe}.

In the TF geometry muons probe the magnetic field distribution $P(B)$ in the sample. In the mixed state of a type II superconductor $P(B)$ is determined by the magnetic penetration depth $\lambda$ and the coherence length $\xi$. The $P(B)$ distributions obtained from the Fourier transform of the $\mu$SR time spectra at $1.6$~K and above $T_c$ are shown in Figs. \ref{fig_ZF_raw}c and e. In the normal state a symmetric $P(B)$ at the position of the applied magnetic field is observed.
The broadening of $P(B)$ in the normal state is due to nuclear and diluted electronic magnetic moments.
Below $T_c$ an additional broadening and an asymmetric line shape $P(B)$ due to the formation of the flux line lattice (FLL) shows up.
The TF $\mu$SR time spectra were analyzed by a theoretical polarization function $A(t)$ by assuming an internal field distribution $P_{\textrm{FLL}}(B)$ and to account for the FLL disorder by multiplying $P_{\textrm{FLL}}(B)$ with a Gaussian function:\cite{Maisuradze,Riseman95}\\
\begin{equation}
A(t)=A_0\, e^{i\phi} e^{-(\sigma_{\rm g}^2+\sigma_{\rm nm}^2)t^2/2-\Lambda_{\rm e} t} \int P_{\textrm{FLL}}(B) e^{i\gamma_\mu B t}dB.
\label{eq_TF}
\end{equation}
Here $A_0$ and $\phi$ are the initial asymmetry and the phase of the muon spin ensemble, respectively, $\sigma_{\rm g}$ is a parameter related to the FLL disorder,\cite{Maisuradze,Riseman95} $\sigma_{\rm nm}$ is the nuclear moment contribution measured at $T>T_c$, which is generally temperature independent,\cite{Schilling82} and $\Lambda_{\rm e}$ is the relaxation rate of the electronic moment contribution, which was obtained from the measurements taken above $T_c$.

The magnetic field distribution $P_{\rm FLL}(B)$ for a FLL of an anisotropic superconductor was determined from the spatial variation of the magnetic field $B(\textbf{r})$ calculated in an orthogonal frame $x$, $y$, $z$ with $H\parallel z$ ($z$ is one of the principal axes $a$, $b$, $c$) using the expression:\cite{Yaouanc97}\\
\begin{equation}
B(\textbf{r})=\langle B \rangle \sum_G \textrm{exp}(-i\textbf{G}\cdot \textbf{r})B_\textbf{G}(\lambda,\xi,b).
\label{field_distribution}
\end{equation}
Here, $\langle B\rangle$ is the average magnetic field in the superconductor (magnetic induction), $b=\langle B\rangle/B_{\rm c2}$ the reduced field ($B_{\rm c2}=\mu_0 H_{\rm c2}$), and $\textbf{r}$ the vector coordinate in a plane perpendicular to the applied field. The Fourier components $B_\textbf{G}$ were obtained within the framework of the Ginzburg-Landau (GL) model.\cite{Yaouanc97}
For a detailed description of the fitting procedure we refer to Ref. \onlinecite{Maisuradze}. The solid lines in Figs. \ref{fig_ZF_raw}c and e correspond to the fast Fourier transforms of the described fits to the $\mu$SR time spectra.

\begin{figure}[t!]
\centering
\vspace{-0.5cm}
\hspace{-1.0cm}
\includegraphics[width=1.1\linewidth]{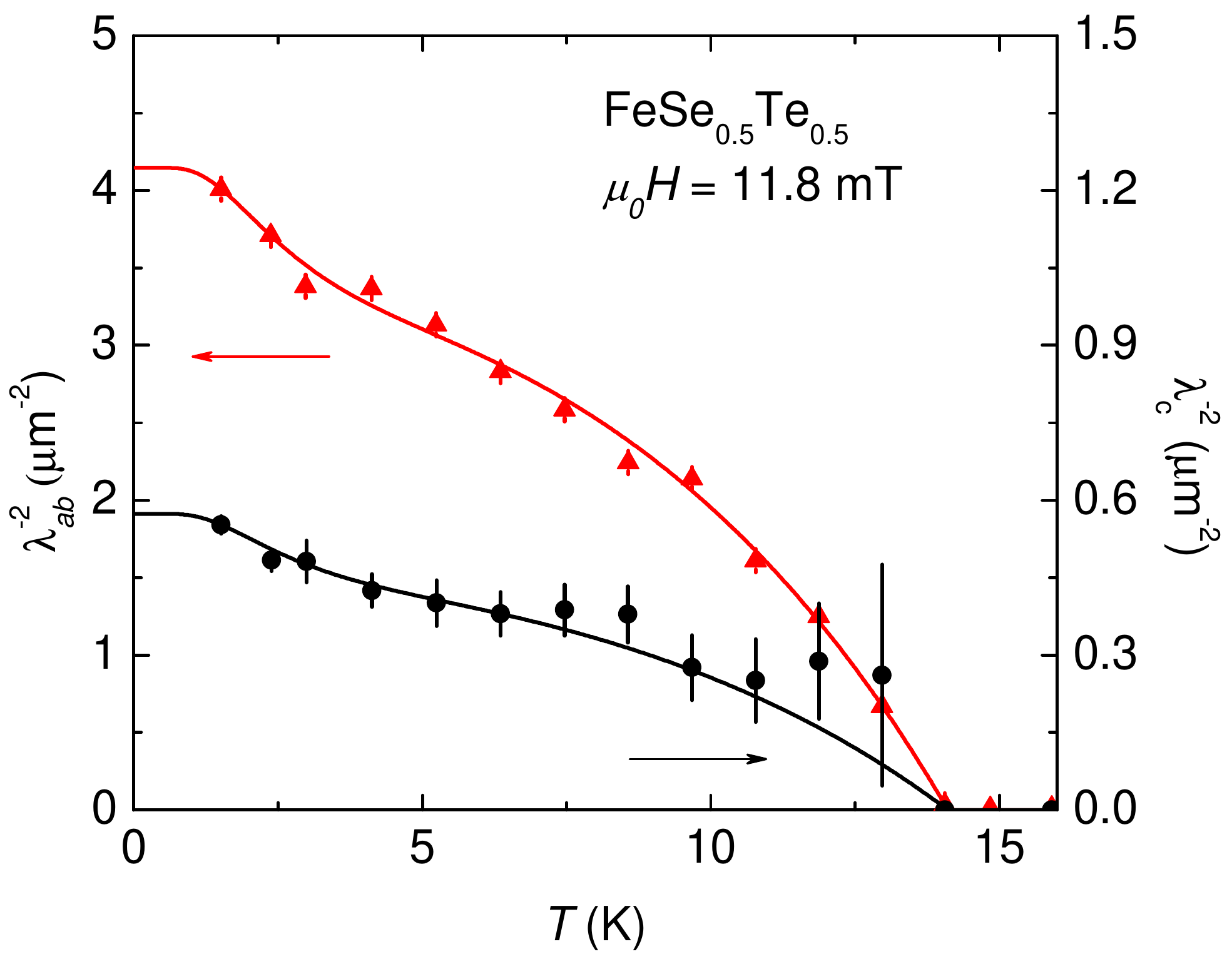}
\caption{(color online) Temperature dependence of the penetration depth components $\lambda_{ab}$ and $\lambda_c$ of single-crystal FeSe$_{0.5}$Te$_{0.5}$. The solid lines correspond to fits using Eq. (\ref{eq_lambda}). The corresponding fit parameters are listed in Table \ref{table_twogap}.}
\label{fig_lambda}
\end{figure}
The temperature dependences of $\lambda^{-2}_{ab}$ and $\lambda^{-2}_c$ extracted from the $\mu$SR time spectra using the fitting procedure described above are shown in Fig. \ref{fig_lambda}.
These data were analyzed within the framework of the phenomenological $\alpha$-model by assuming that $\lambda^{-2}$ is a linear combination of two terms:\cite{Carrington03}
\begin{equation}
\frac{\lambda^{-2}(T)}{\lambda^{-2}(0)}=w\frac{\lambda^{-2}(T,\Delta_S^0)}{\lambda^{-2}(0,\Delta_S^0)}+(1-w)\frac{\lambda^{-2}(T,\Delta_L^0)}{\lambda^{-2}(0,\Delta_L^0)}.
\label{eq_lambda}
\end{equation}
Here, $\Delta_S^0$ and $\Delta_L^0$ are the zero-temperature values of the small and the large gap, respectively, and $w$ ($0\leq w\leq 1$) is the weighting factor which measures the relative contribution of the two gaps to $\lambda^{-2}(T)/\lambda^{-2}(0)$. For the temperature dependence of $\lambda^{-2}$ of a London superconductor ($\lambda \gg \xi$) with a $s$-wave gap the following relation can be used:\cite{Tinkham}
\begin{equation}
\frac{\lambda^{-2}(T,\Delta_{S(L)}^0)}{\lambda^{-2}(0,\Delta_{S(L)}^0)}=1+2\int_{\Delta(T)}^{\infty}\left( \frac{\partial f}{\partial E} \right)\frac{E}{\sqrt{E^2-\Delta^2(T)}}dE.
\end{equation}
Here $\lambda(0)$ is the zero temperature value of the magnetic penetration depth, $f(E)=[1+\exp(E/k_BT)]^{-1}$ is the Fermi function ($E$ is the excitation energy, $k_B$ is the Boltzmann constant), and $\Delta(T)=\Delta(0)\tilde{\Delta}(T/T_c)$ represents the temperature dependence of the gap with $\tilde{\Delta}(T/T_c)=\tanh(1.82[1.018(T_c/T-1)^{0.51}])$.\cite{Carrington03}\\
\begin{table}[t!]\centering\caption{Summary of the parameters obtained for single-crystal FeSe$_{0.5}$Te$_{0.5}$ by means of $\mu$SR and magnetization measurements. The errors of the $\mu$SR data are statistical errors and do not take into account any systematical errors that may be present in the data.}\hspace{20mm}
\begin{tabular}{l|c c|c c}\toprule
 &\multicolumn{2}{c|}{$\mu$SR}&\multicolumn{2}{c}{magnetization}\\
 & $ab$-plane & $c$-axis      & $ab$-plane & $c$-axis\\\hline
$T_c$~(K)&\multicolumn{2}{c|}{14.1(1)}&\multicolumn{2}{c}{14.6(1)}\\
$\Delta_S$~(meV)&\multicolumn{2}{c|}{0.51(3)}&\multicolumn{2}{c}{-}\\
$2\Delta_S/k_BT_c$&\multicolumn{2}{c|}{0.84(5)}&\multicolumn{2}{c}{-}\\
$\Delta_L$~(meV)&\multicolumn{2}{c|}{2.61(9)}&\multicolumn{2}{c}{-}\\
$2\Delta_L/k_BT_c$&\multicolumn{2}{c|}{4.3(1)}&\multicolumn{2}{c}{-}\\
$w$&0.32(1)&0.36(2)&\multicolumn{2}{c}{-}\\
$\lambda_{ab,\;c}(0)$~(nm)&491(8)&1320(14)&460(100)&1100(300)\\
$H_{\rm c1}$~(mT)&\multicolumn{2}{c|}{-}&2.0(2)&4.5(3)\\\toprule
\end{tabular}
\label{table_twogap}
\end{table}
The temperature dependences of $\lambda_{ab}$ and $\lambda_c$ were determined simultaneously, assuming the same values for the small and large gap ($\Delta_{S,ab}=\Delta_{S,c}$ and $\Delta_{L,ab}=\Delta_{L,c}$), but different weighting factors $w$. The results of this analysis are summarized in Table \ref{table_twogap}.
The ratios $2\Delta_S/k_BT_c=0.84(4)$ and $2\Delta_L/k_BT_c=4.3(1)$ are close to what was reported for isostructural FeSe$_{1-x}$\cite{khasanovFeSe}.
Based on scanning tunneling spectroscopy measurements, Kato {\it et al.} \cite{kato09} reported for FeSe$_{0.4}$Te$_{0.6}$ only one $s$-wave gap $\Delta \simeq 2.3$~meV. This value is quite similar to our result of the large gap ($\Delta_{L}=2.61(9)$~meV). However, a single $s$-wave gap is not sufficient to describe the present $\mu$SR data. The weighting factors $w$ are about the same for $1/\lambda_{ab}^2$ and $1/\lambda_c^2$. Similar results were already reported for isostructural FeSe$_{1-x}$.\cite{khasanovFeSe}
Recently, Kim {\it et al.}\cite{Kim10} reported on magnetic penetration depth measurements on Fe$_{1.03}$Se$_{0.37}$Te$_{0.63}$ by means of a radio-frequency tunnel diode resonator technique. Their value $\lambda_{ab}(0)\simeq 560(20)$~nm is in good agreement with the value reported here (see Table \ref{table_twogap}). Furthermore, they found a clear signature of multi-gap superconductivity with compareable gap values ($\Delta_S\simeq 1.2$~meV and $\Delta_L\simeq 2$~meV).
In a recent $\mu$SR study of polycrystalline FeSe$_{0.5}$Te$_{0.5}$ the temperature dependence of $\lambda_{ab}$ was found to be compatible with either a two gap $s$+$s$-wave or anisotropic $s$-wave model with $\lambda_{ab}=534(2)$~nm.\cite{Biswas10}
For the $s$+$s$-wave analysis the following results were obtained: $\Delta_{L}(0)=2.6(1)$~meV, $\Delta_{S}(0)=0.87(6)$~meV, and $1-w=0.70(3)$.\cite{Biswas10}
These results are in fair agreement with the present results listed in Table \ref{table_twogap}.

\begin{figure}[t!]
\centering
\vspace{-0cm}
\includegraphics[width=1\linewidth]{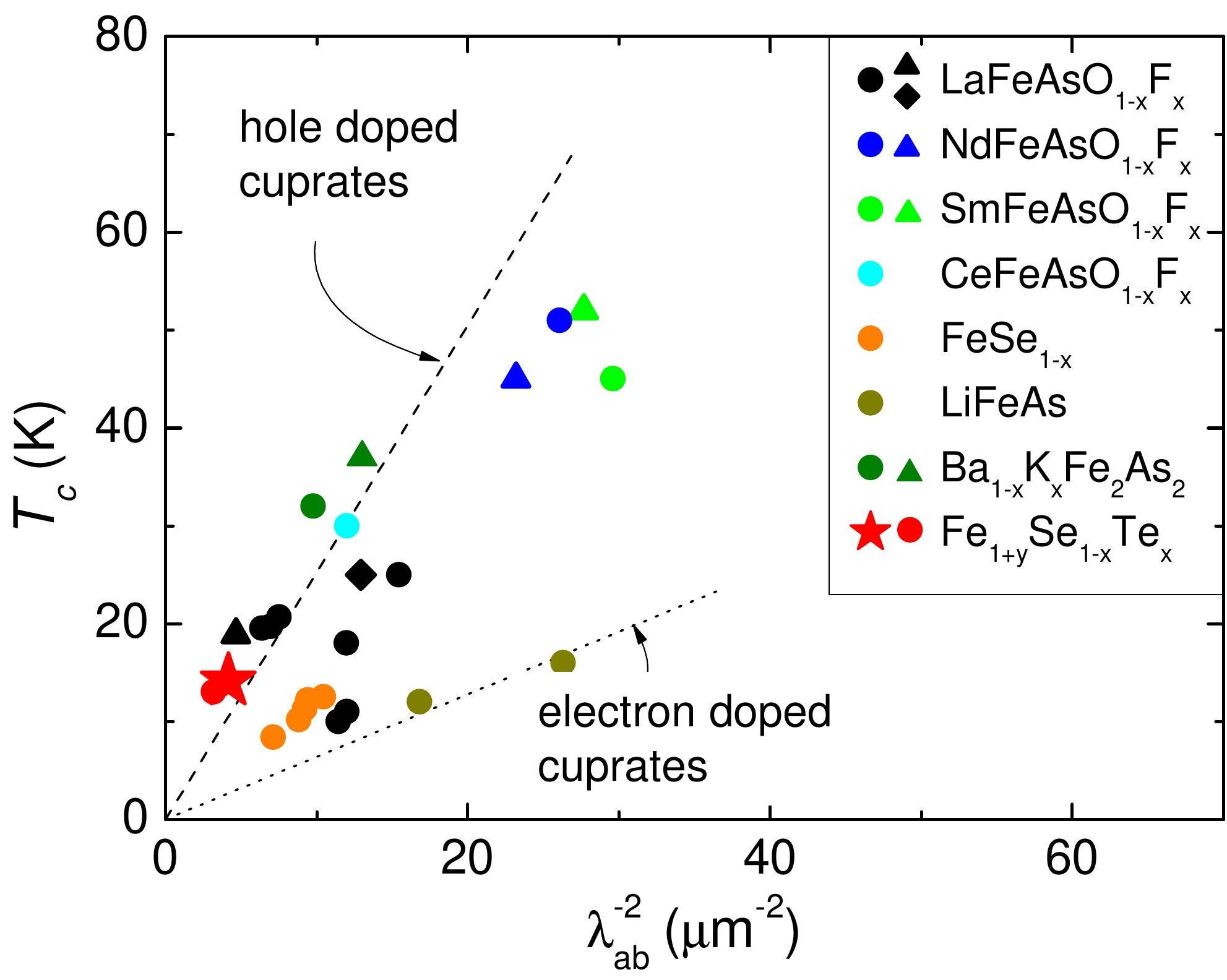}
\caption{(color online) Uemura plot for a selection of some Fe-based HTS. The Uemura relation observed for underdoped cuprates is included for comparison as a dashed line for hole doping and as a dotted line for electron doping (after Ref. \onlinecite{Shengelaya05}). LaFeAsO$_{1-x}$F$_x$ data from Refs. \onlinecite{Luetkens09,Luetkens09_PRL} ($\bullet$), Ref. \onlinecite{Takeshita09} ($\blacktriangle$), and Ref. \onlinecite{Carlo09} ($\blacklozenge$); NdFeAsO$_{1-x}$F$_x$ data from Ref. \onlinecite{Khasanov_Sm} (\textbullet) and Ref. \onlinecite{Carlo09} ($\blacktriangle$); SmFeAsO$_{1-x}$F$_x$ data from Ref. \onlinecite{Khasanov_Sm} (\textbullet) and Ref. \onlinecite{Drew09} ($\blacktriangle$); CeFeAsO$_{1-x}$F$_x$ data from Ref. \onlinecite{Carlo09}, FeSe$_{1-x}$ data from Refs. \onlinecite{Khasanov_FeSepressure,khasanovFeSe}, LiFeAs data from Ref. \onlinecite{Pratt09}, Ba$_{1-x}$K$_x$FeAs data from Ref. \onlinecite{Khasanov09_Ba}, Fe$_{1+y}$Se$_{1-x}$Te$_x$ data from Ref. \onlinecite{Kim10} (\textbullet). The red star ($\bigstar$) is showing the data for FeSe$_{0.5}$Te$_{0.5}$ obtained in this work. }
\label{fig_uemura}
\end{figure}

Uemura \emph{et al.}\cite{Uemura89} found an empirical relation between the zero temperature superfluid density $\rho_s(0)\propto\lambda_{ab}^{-2}(0)$ and $T_c$ which seems to be generic for various families of cuprate HTS (Uemura plot). This ``universal'' relation $T_c(\rho_s)$ has the following features: With increasing carrier doping $T_c$ initially increases linearly ($T_c\propto \rho_s(0)$), then saturates, and finally is suppressed for high carrier doping. It is interesting to check whether the Uemura relation also holds for iron-based superconductors. For this reason $T_c$ vs. $\lambda_{ab}^{-2}(0)$ is plotted in Fig. \ref{fig_uemura} for a selection of various Fe-based superconductors investigated so far.\cite{Luetkens09,Luetkens09_PRL,Takeshita09,Carlo09,Khasanov_Sm,Drew09,Khasanov_FeSepressure,khasanovFeSe,Pratt09,Khasanov09_Ba,Kim10} For comparison the linear parts of the Uemura relation for hole-doped (dashed line) and electron-doped (dotted line) cuprate HTS are also shown in Fig. \ref{fig_uemura}. Due to the small number of data points available for a particular family of Fe-based superconductors there is no obvious trend visible. However, all data points are located within an area determined by the straight lines representing the hole-doped and electron-doped cuprates. Whereas various of the Fe-based HTS, including FeSe$_{0.5}$Te$_{0.5}$ (red star in Fig. \ref{fig_uemura}) investigated here, are located near the hole-doped cuprates in the Uemura plot, the ``111'' system appears to be close to the electron-doped cuprates.

\section{Temperature dependent anisotropy parameters}
For a conventional single-band $s$-wave layered superconductor the anisotropy parameter is defined as:\cite{Tinkham}
\begin{equation}
\gamma=\sqrt{m_c^*/m_{ab}^*}=\lambda_{c}/\lambda_{ab}=H_{\rm c2}^{\parallel ab}/H_{\rm c2}^{\parallel c}=\xi_{ab}/\xi_{c}.
\label{eq_anisotropy}
\end{equation}
Here, $m_{ab}^*$ and $m_c^*$ are the effective charge carrier masses related to supercurrents flowing in the $ab$-planes and along the $c$-axis, respectively.
Whereas the cuprates were characterized by a well-defined effective mass anisotropy, the observation of two distinct anisotropy parameters in MgB$_2$ challenged the understanding of anisotropic superconductors.\cite{Angst02,Angst04,Fletcher05}
Various experiments, such as magnetic torque,\cite{Weyeneth09,Weyeneth09a} tunneling,\cite{Gonnelli09,Samuely09} point contact and infrared spectroscopy,\cite{Szabo09,Li08} as well as the measurements of the specific heat,\cite{Mu09} the lower and upper critical field,\cite{Ren08_PRL,Hunte08} and the superfluid density\cite{Weyeneth10,Malone09,Hiraishi09,Khasanov09_Sr,Khasanov09_Ba,khasanovFeSe} indicate that Fe-based pnictides are multi-gap superconductors having unconventional anisotropic properties,\cite{Weyeneth09a,Weyeneth09,Balicas09} similar to MgB$_2$.\cite{Nagamatsu01,Gonelli02}

\begin{figure}[t!]
\centering
\vspace{-0cm}
\includegraphics[width=1\linewidth]{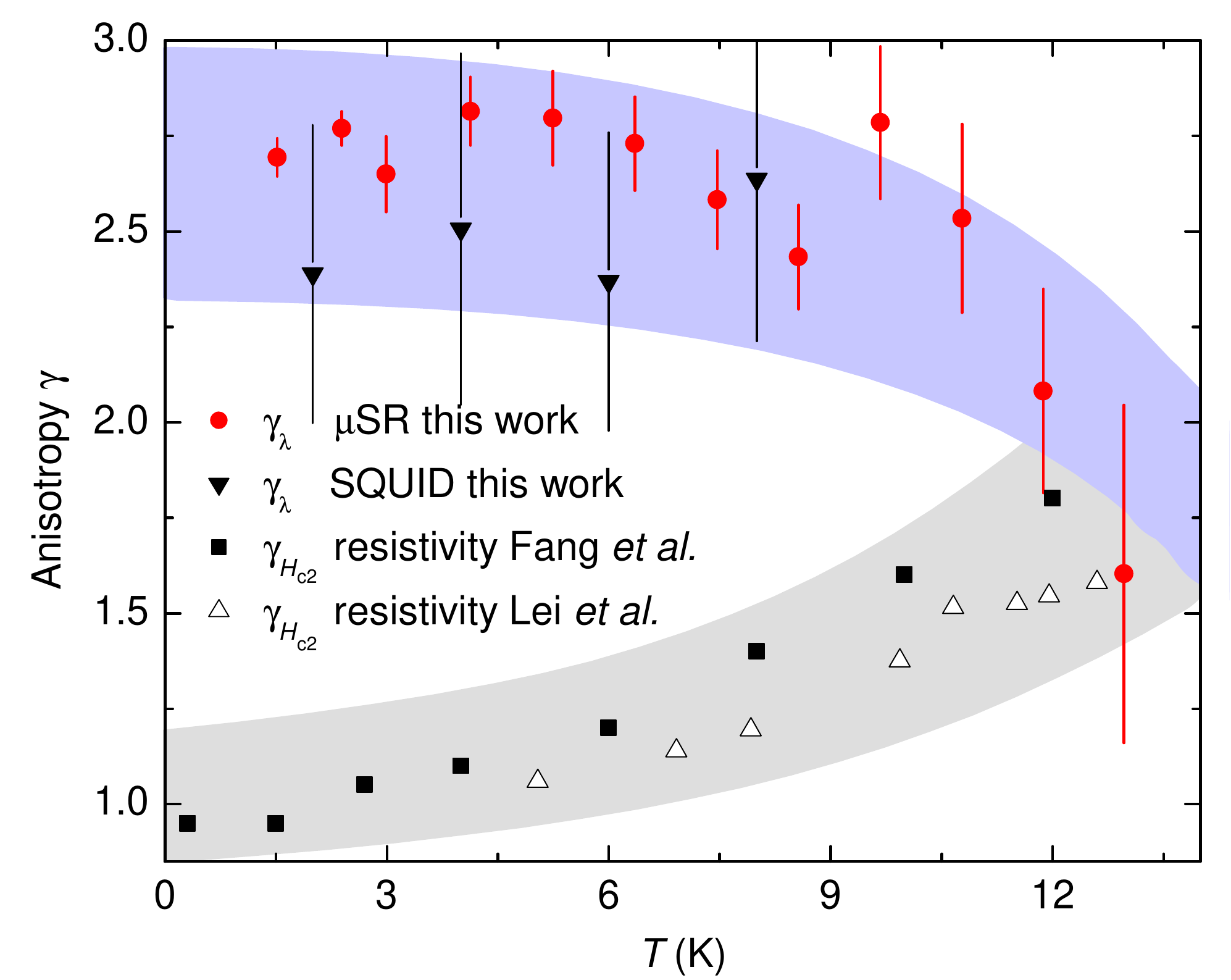}
\caption{(color online) Comparison of the temperature dependence of the magnetic penetration depth anisotropy parameter $\gamma_\lambda=\lambda_{c}/\lambda_{ab}$ measured by $\mu$SR and by SQUID for single-crystal FeSe$_{0.5}$Te$_{0.5}$ with the $H_{\rm c2}$-anisotropy parameter $\gamma_{H_{\rm c2}}=H_{c2}^{\parallel ab}/H_{c2}^{\parallel c}$ obtained from resistivity measurements for Fe$_{1.11}$Se$_{0.4}$Te$_{0.6}$ by Fang {\it et al.} \cite{fang09} and for Fe$_{1.02}$Se$_{0.39}$Te$_{0.61}$ by Lei {\it et al.}\cite{Lei10}. The lines are guides to the eyes. }
\label{fig_anisotropy}
\end{figure}

The temperature dependence of the magnetic penetration depth anisotropy parameter $\gamma_\lambda = \lambda_c/\lambda_{ab}$ extracted from the $\mu$SR data (see Fig. \ref{fig_lambda}) is shown in Fig. \ref{fig_anisotropy}. Note that $\gamma_{\lambda}$ increases with decreasing temperature and saturates at $\gamma_\lambda\simeq2.6(3)$ at low temperatures. This observation is further supported by the temperature dependence of $\gamma_\lambda$ determined from the lower critical field measurements presented in Fig. \ref{fig_Hc1_T}. In this case $\gamma_\lambda$ is readily obtained from Eqs. (\ref{eq_Hc1a}) and (\ref{eq_Hc1b}):\cite{Tinkham}
\begin{equation}
\gamma_\lambda=\frac{\lambda_{c}}{\lambda_{ab}}=\frac{H_{{\rm c}1}^{||c}}{H_{{\rm c}1}^{||ab}}\left(1+\frac{\ln(\gamma_\lambda)+\ln(\gamma_{H_{\rm c2}})}{2\ln(\kappa_{ab})+1}\right)
\label{eq_lambdaHc1}
\end{equation}
Here, $\kappa_{ab}=\lambda_{ab}/\xi_{ab}$ denotes the Ginzburg-Landau parameter. In this work $\kappa_{ab}$ was estimated to be $\simeq180$ from present experiments.\cite{fang09} The values of $\gamma_\lambda$ extracted from the SQUID data using Eq. (\ref{eq_lambdaHc1}) are also depicted in Fig. \ref{fig_anisotropy} and are in fair agreement with those obtained from the $\mu$SR data.

The upper critical field anisotropy parameter, $\gamma_{H_{\rm c2}}= H_{\rm c2}^{\parallel ab}/H_{\rm c2}^{\parallel c}=\xi_{ab}/\xi_{c}$, was studied by Fang {\it et al.}\cite{fang09} and Lei {\it et al.}\cite{Lei10} by resistivity measurements on Fe$_{1+y}$Se$_{0.4}$Te$_{0.6}$ ($y=0.02$ and 0.11).
These data are plotted in Fig. \ref{fig_anisotropy} as well.
Note that $\gamma_{H_{\rm c2}}$ decreases with decreasing temperature.
Obviously, the behavior of the two distinct anisotropy parameters $\gamma_{\lambda}$ and $\gamma_{H_{\rm c2}}$ is {\it not} consistent with Eq. (\ref{eq_anisotropy}).
The observed behavior is similar to the one of the two-gap superconductor MgB$_2$ and other Fe-based superconductors.\cite{Weyeneth09}
For MgB$_2$, however, $\gamma_{\lambda}$ decreases with decreasing temperature while $\gamma_{H_{\rm c2}}$ increases.\cite{Angst02}

\section{Conclusions}

Single crystals with a nominal composition of FeSe$_{0.5}$Te$_{0.5}$ were studied by means of muon spin rotation ($\mu$SR), SQUID and torque magnetometry, and neutron powder diffraction. At room temperature the crystal shows mainly a tetragonal phase of PbO type that becomes orthorombic and superconducting at low temperatures. The stoichiometry was refined to Fe$_{1.045}$Se$_{0.406}$Te$_{0.594}$. The onset transition temperature is $T_c=14.6$~K, and the lower critical field values measured for both crystallographic directions were determined at zero temperature as $H^{\parallel ab}_{\rm c1}(0)=2.0(2)$~mT and $H^{\parallel c}_{\rm c1}(0)=4.5(3)$~mT.

By means of $\mu$SR it was found that for FeSe$_{0.5}$Te$_{0.5}$ the temperature dependence of the magnetic penetration depth for both crystallographic directions is best described by a two gap $s$+$s$-wave model with zero-temperature values of the magnetic penetration depth of $\lambda_{ab}(0)=491(8)$~nm and $\lambda_{c}(0)=1320(14)$~nm, consistent with recent $\mu$SR results obtained for a polycrystalline sample.\cite{Biswas10}
This two-gap scenario is in line with the generally accepted view of multi-gap superconductivity in Fe-based HTS.
Evtushinsky {\it et al.}\cite{Evtushinsky09} pointed out that most Fe-based HTS exhibit two gaps, a large one with $2\Delta/k_BT_c=7(2)$ and a small one with $2.5(1.5)$.
The magnitudes of the large and the small gap for FeSe$_{0.5}$Te$_{0.5}$ ($2\Delta_{S}/k_BT_c=0.84(4)$ and $2\Delta_{L}/k_BT_c=4.3(1)$) are at the lower limit for Fe-based HTS.
Moreover, the magnetic penetration depth anisotropy parameter $\gamma_\lambda$ determined from penetration depth experiments by means of $\mu$SR, is within experimental error the same as the one deduced from $H_{\rm c1}$ measurements. Both techniques yield a temperature dependent $\gamma_\lambda$ that increases with decreasing temperature from 1.6 at $T_c=14.6$~K to 2.6 at $T=1.6$~K.
Compared to SmFeAsO$_{0.8}$F$_{0.2}$ and NdFeAsO$_{0.8}$F$_{0.2}$,\cite{Weyeneth09} superconducting FeSe$_{0.5}$Te$_{0.5}$ is much more isotropic, but quite compareable to the 122 class of Fe-based superconductors.\cite{Bukowski09,Khasanov09_Sr,Khasanov09_Ba}
This suggests that the direct electronic coupling of the Fe$_{2}$Se$_{2}$ layers in the "11" system is similar to the one through the intervening $Ae$-layers in the "122" class of superconductors, but more effective than the coupling through the $Ln$O layers in the "1111" Fe-based systems.
While $\gamma_\lambda$ increases with decreasing temperature the anisotropy parameter of the upper critical field $\gamma_{H_{\rm c2}}$ determined by resistivity measurements decreases.\cite{fang09,Lei10}
The observed behavior is similar to that of the two-gap superconductor MgB$_2$ and other Fe-based superconductros and supports a two-gap scenario also in FeSe$_{0.5}$Te$_{0.5}$.\cite{Weyeneth09}
Note, however, that for MgB$_2$ the slopes of $\gamma_{\lambda}(T)$ and $\gamma_{H_{\rm c2}}(T)$ have reversed signs\cite{Angst02,Fletcher05} as compared to the Fe-based superconductors. The reason for this difference is still unclear.
Furthermore, the value of $\lambda^{-2}_{ab}(0)$ for FeSe$_{0.5}$Te$_{0.5}$ extracted from $\mu$SR data as well as the values of $\lambda^{-2}_{ab}(0)$ obtained for various Fe-based superconductors fall on the Uemura plot\cite{Uemura89} within the limits of hole-doped and electron-doped cuprates.\cite{Shengelaya05} This suggests that the pairing mechanism in the Fe-based superconductors is unconventional, as is also the case for the cuprates.

In conclusion, FeSe$_{0.5}$Te$_{0.5}$ shows evidence for two-gap superconductivity, which is reflected in the temperature dependence of $\lambda^{-2}$ and by the existance of two distinct anisotropy parameters $\gamma_\lambda(T)$ and $\gamma_{H_{\rm c2}}(T)$. The two-gap scenario is observed for most Fe-based superconductors, suggesting that this behavior is generic for layered high-temperature superconductors: It is strongly supported by various experiments for Fe-based superconductors (Ref. \onlinecite{Evtushinsky09} and references therein), it is well established for MgB$_2$,\cite{Angst02,Angst04} and there is firm evidence for two-gap superconductivity also in the cuprates.\cite{Khasanov07,Bussmann,Khasanov_JofSupercond,Khasanov07_1} However, it remains to be seen whether superconductivity in these classes of high-temperature superconductors has the same or a similar origin.

\section{Acknowledgments}
The $\mu$SR experiments were performed at the Swiss Muon Source, Paul Scherrer Institut, Villigen, Switzerland. This work was partially supported by the Swiss National Science Foundation, the EU Project CoMePhS, the Polish Ministry of Science and Higher Education with the research project No. N N202 4132 33, and by the NCCR Program MaNEP.

\end{document}